\documentclass[
    ,final            
  ]
  {aipproc}

\layoutstyle{6x9}

\newcommand{\etal}{\mbox{et al.}}

\newcommand{\msun}{M$_\odot$}

\newcommand{\chandra}{{\it Chandra}}
\newcommand{\xmm}{{\it XMM}}

\newcommand{\apj}{{\it ApJ}}

\newcommand{\apjs}{{\it ApJS}}

\newcommand{\aap}{{\it A\&A}}
\newcommand{\aaps}{{\it A\&AS}}

\newcommand{\mnras}{{\it MNRAS}}

\newcommand{\nat}{{\it Nature}}

\newcommand{\ergs}{\mbox{erg s$^{-1}$}}

\newcommand{\arcmin}{$^\prime$}
\newcommand{\degree}{$^\circ$}

\newcommand{\wdpulsar}{\mbox{CXOU J164710.2-455216}}

\begin{document}

\title{Which Stars Form Black Holes and Neutron Stars?}


\author{Michael P. Muno}{
  address={Space Radiation Laboratory, California Institute of Technology, Pasadena, CA 91125}
}

\begin{abstract}
I describe the current state of our knowledge of the mapping
between the initial masses of stars and the compact objects --- 
particularly neutron stars and black holes --- that they produce. 
Most of that knowledge is theoretical in nature, and relies on 
uncertain assumptions about mass loss through winds, binary mass 
transfer, and the amount of mass ejected during a supernovae. Observational
constraints on the initial masses of stars that produce neutron
stars and black holes is scarce. They fall into three general categories: 
(1) models of the stars that produced the supernova remnants associated 
with known compact objects, (2) scenarios through with high mass X-ray
binaries were produced, and (3) associations between compact objects
and coeval clusters of stars for which the minimum masses of stars 
that have undergone supernovae are known. I focus
on the last category as the most promising in the near term. I describe
three highly-magnetized neutron stars that have been 
associated with progenitors that had initial masses of $>$30\msun,
and evaluate the prospects of finding further associations between 
star clusters and compact objects.
\end{abstract}

\maketitle


\section{Introduction}

In introductory astronomy classes, the fates of stars are generally outlined
as follows: stars with initial masses $M_i$$<$8 \msun\ never burn all their
nuclear fuel, and eventually become white dwarfs; stars 
with 8$<$$M_i$$<$25 \msun\ produce supernovae that eject most of the mass
of the star, and leave behind neutron stars; and stars with $M_i$$>$25 \msun\ 
are so massive that during supernovae, much of the mass remains bound to 
the core, so it falls back and causes them to collapse into black holes. What 
is not generally appreciated is that this story is based almost entirely on 
theoretical calculations that contain several uncertain elements; 
observational data is scarce \citep{heg03}. Establishing observational 
constraints on 
these calculations is important because supernovae are the largest source 
of mass and energy input into the interstellar medium \citep{lrd92}. 
Therefore, knowing whether a star will form a neutron star or a
black hole affects our understanding of how rapidly galaxies can 
assemble \citep{sca06}, how the abundances of metals in galaxies increase
with time \citep{tsu95}, and the amount of warm matter ejected into the 
intergalactic medium \citep{del04}. 

There are several reasons that predicting whether a massive star will
produce a neutron star or a black hole is difficult.  First, massive
stars lose a significant amount of mass through stellar winds as they
evolve, but the rates are still uncertain by factors of several
\citep{ll93,blh06}.  This problem is most acute for massive
stars that become luminous blue variables, because their evolutionary
models do not yet include the possibility that they can lose $\sim$10
\msun\ of mass in eruptions like that seen from eta Carina in the
late 19th century \citep{so06}.  Second, a large, but
poorly-determined, fraction of massive stars are in binaries. Mass
transfer between the binary stars can occur as they evolve, which
would in turn change the mass of each star when it undergoes a
supernova \citep{wl99}. Third, the exact mechanism that turns a
collapsing stellar core into a supernova is still uncertain. In
particular, it is unknown to what degree rapid rotation and strong
magnetic fields might be increase the amount of mass lost during the
supernova explosion \citep{tcq05,aw05}, and consequently the mass left 
behind in a compact remnant. 

One way in which the complexity of this problem is manifest from an
observational standpoint is in the zoology of neutron stars \citep{ptp06}. 
It was once assumed that all neutron stars were born as rapidly-rotating
radio pulsars, which are powered by particles that are accelerated in their 
$B$$\sim$$10^{12}$ G fields as they spin down. However, it now appears the 
neutron stars are
born with a variety of field strengths and rotation rates. First, a class 
of isolated X-ray pulsars, consisting of soft gamma repeaters and anomalous 
X-ray pulsars, appear to be magnetars that are powered by the decay of 
$>$$10^{14}$ G fields \citep{wt06}. Second, it has been suggested that 
the planets around the isolated millisecond pulsar PSR 1257$+$12 could
only have survived if that neutron star was formed in something like its
present state, rapdily rotating but with with a weak magnetic field 
($B$$\sim$$10^{9}$ G) \citep{mh01}. Finally, several manifestations of 
neutron stars have not been definitively associated with the above 
categories, including the central compact objects in supernova remnants 
\citep{cha01,sew03,hb06}, the ``Magnificent Seven'' isolated cooling 
neutron stars \citep{wwn96,hab06}, and the rotating radio transients
\citep{mcl06}.  Understanding how this diversity arises should provide
insight into how stellar mass, rotation, and magnetic fields affect
the mass and energy ejected during a supernovae.



\section{Observational Constraints}

In principle, there are several ways in which one could determine the 
initial masses of stars that produced black holes and neutron stars:  
by modeling the supernova remnants associated with compact objects, 
by creating scenarios by which high-mass X-ray binaries (HMXBs) 
formed, and by identifying compact objects in coeval star clusters.
Each of these has their own strengths and complications.

In principle, if one has identified a supernova associated with a known
compact object, it should be possible to infer the mass of the 
star that exploded from the masses and abundances of metals in the remnant.
The Crab nebula is best modeled as descending from an 8--10 \msun\ progenitor, 
\citep{nom82}\footnote{The more massive progenitor suggested by 
\citet{mac89} is now disfavored \citep[e.g.,][]{mac96}.}
so it is not surprising that it contains a neutron star in the form
of a radio pulsar. Likewise, the abundances in the remnant G292.0+1.8 
generally matched those produced by a 25 \msun\ progenitor in some models 
\citep{hs94,thi96}. This is at the upper end of the boundary at which a 
star is often considered likely to produce a neutron star, and indeed
a radio pulsar is associated with the remnant \citep{cam02,hug03}.
However, Cas A has been variously modeled as originating from a 
20--25 \msun\ single progenitor \citep{lh03}, or from a
15--25 \msun\ star that was in a binary \citep{you06}. The models are 
complex and do not yet match all of the available data on supernova remnants.
Nonetheless, this technique is promising, because 15 of the 45 supernovae
within 5 kpc of Earth are associated with radio pulsars or X-ray
emitting central compact objects \citep{kap04}.

The second method is to generate scenarios through which HMXBs could 
have formed. More massive stars end their 
lives first, so one would tend to expect that the mass of the mass donor 
in the binary would represent a lower limit to the mass of the progenitor
to the compact object \citep{erg98}. Cyg X-1 is the prototypical black 
hole X-ray binary, and provides striking evidence that black holes 
can indeed form --- it contains a 16 \msun\ compact object accreting
matter from a 33 \msun\ O9 supergiant \citep{gb86}.
Moreover, GX 301-2 is a neutron star X-ray pulsar in a binary with 
the 50 \msun\ B hypergiant Wray 977 \citep{kap95}. This could suggest that  
the most massive stars sometimes form neutron stars instead of black holes 
\citep{erg98}. However, \citet{wl99} have constructed scenarios in 
which the neutron star GX 301-2 formed from a $25$\msun\ star that 
transferred most of its mass to its companion. Between these results,
4U 1700--37 contains a 50--60 \msun\ 06.5 supergiant in a binary with
a 2.4 \msun\ compact object of undetermined nature \citep{cla02}.
This technique must be pursued further, because although the uncertainties 
in how mass transfer proceeds are formidable, HMXBs provide the 
only constraints on the birth masses of compact objects.

The third method is to identify compact objects that are members of 
coeval populations of stars, for which the maximum initial masses
of stars still present is also a lower limit on the masses of stars
that have undergone supernovae. Unfortunately, neutron stars and 
black holes may receive kicks at birth \citep{mir02,hob05}, so there 
is only a short window of time in which one can expect to find a 
compact object still associated with a star cluster. In rare cases, it 
may be possible to trace the three-dimensional motions of compact 
objects back to their birth clusters. This has been attempted for 
Geminga, for which the proper motion was measured, and the line-of-sight
velocity was estimated by modeling the bow shock produced by the
pulsar wind nebulae. It was suggested that Geminga
formed in the Cas-Tau OB or Ori OB 1a associations, which would imply
that its progenitor was $<$15 \msun\ \citep{pel05}. 

Fortunately, in recent years several more straightforward examples
have emerged: three magnetars have been found to 
reside in massive young star clusters \citep{fuc99,vrb00,mun06}. These 
provide unambiguous lower limits on the masses of progenitors 
neutron stars, and the results are surprising. I describe these below,
focusing on the most clear result of the group, the identification of
a slow X-ray pulsar in the Galactic star cluster Westerlund 1.

\begin{figure}
\includegraphics[width=7.0cm]{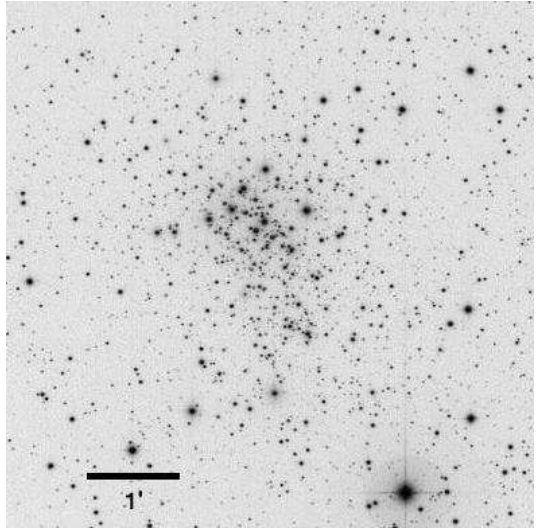}
\caption{
I band Image of Westerlund 1 taken with the ESO 2.2m MPG+WFI (copyright
ESO). The faintest cluster members in this image are O7 V stars 
\citep{cla05}.}
\end{figure}

\begin{figure}
\includegraphics[width=7.25cm]{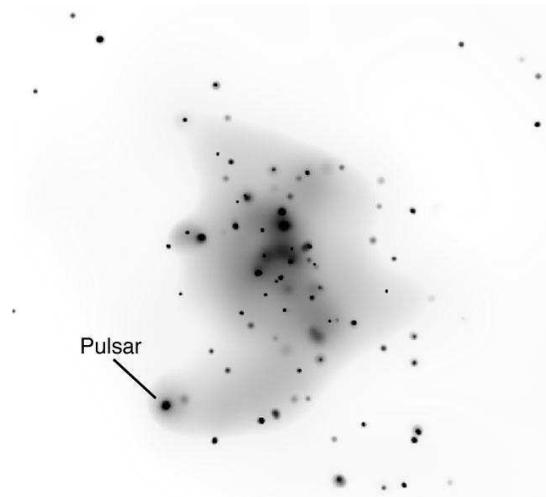}
\caption{\chandra\ image of Westerlund 1, with
the location of the 10.6 s X-ray pulsar marked \citep{mun06}.}
\end{figure}


\section{Magnetars in Star Clusters}

Among the star clusters associated with magnetars, Westerlund 1 is 
the most well-studied (Fig. 1). The cluster contains 24 Wolf-Rayet stars, 
4 red supergiant, 4 yellow hypergiants, over 80 OB supergiants, and
numerous main-sequence stars as early as O7 \citep{wes87,cla05}. 
The cluster contains $\sim$$10^{5}$ \msun\ of stars in a region 6 pc across
($D$=5 kpc), and is coeval with an age of 4--5 Myr \citep{cro06}.
It is therefore just old enough for $>$40 \msun\ stars
to have undergone supernovae, with a rate of approximately 1 per
10,000 years.

A group of us observed the cluster with \chandra\ hoping to find black
hole HMXBs, but instead we found a 10.6 s X-ray pulsar, \wdpulsar\
(Fig. 2) \citep{mun06,ski06}. Although we have not yet measured the spin-down
rate for the pulsar ($\dot{P} < 9\times10^{-12}$ s s$^{-1}$) 
\citep{mun06b,gav06}, it
exhibits all of the other properties of a magnetar
\citep{wt06}. First, its persistent luminosity is $3\times10^{33}$
\ergs\ (0.5--8.0 keV), which is on the faint end of the range for
magnetars. However, the X-ray luminosity is still larger than the
upper limit to the spin-down power, so it is not a radio pulsar. 
Second, the spectrum of \wdpulsar\
in quiescence can be modeled as a 0.6 keV blackbody originating from
0.1\% of the stellar surface, which is inconsistent with the emission
from an ordinary ($B$$\sim$$10^{12}$ G) cooling neutron star. Third,
there is no infrared counterpart down to $K$$>$18.5, which precludes
its membership in a binary with a companion more massive than a 1
\msun\ pre-main-sequence star. Finally, and most conclusively, the
Burst Alert Telescope on Swift recently detected a soft gamma-ray
burst from \wdpulsar\ with a fluence of $10^{38}$ erg and a duration of 
20 ms \citep{kri06}. This was followed by an increase in the persistent
luminosity of the pulsar by a factor of 100, to 
$1\times10^{35}$ \ergs\ (0.5--8.0 keV) \citep{ci06}. 
Such an outburst is a hallmark of a magnetar \citep{gav02, woo04}.

\wdpulsar\ is almost certainly member of Westerlund 
1. It is located only 1.5\arcmin\ from the center of the cluster,
and based on other observations of the Galactic plane we would expect
that $<$10\% of X-ray sources at this offset are random interlopers 
(Fig. 3). Moreover, through another collaboration (A. Nechita, B. Gaensler, 
\etal, in prep.), we have searched 300 archival \chandra\ and \xmm\ 
observations of the Galactic plane ($|b| < 5$\degree) for new X-ray 
pulsars, and found no new examples with periods between 5--30 s. If 
we use this to establish an upper limit to the surface density of
magnetars toward the Galactic plane, we find that there is a $<$0.5\%
chance of finding a new example in any given \chandra\ observation.
Therefore, with 99.95\% confidence, \wdpulsar\ is a member of
Westerlund 1, and the progenitor of the pulsar had an
initial mass of $>$40 \msun.

\begin{figure}
\includegraphics[width=9.0cm]{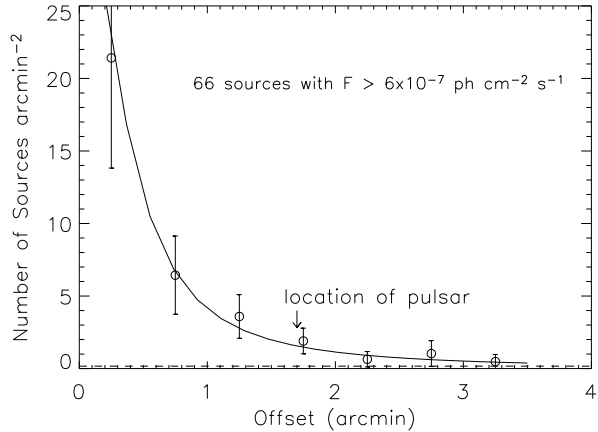}
\caption{
The radial distribution of X-ray sources in Westerlund 1. The 
dashed line denotes the expected surface density of sources in 
the Galactic plane \citep{mun06}.}
\label{fig:dist}
\end{figure}

Similar results have been obtained for three other magnetars. 
SGR 1806-20 is a member of a star cluster containing at least
4 carbon-type Wolf Rayet stars, 3 OB supergiants, and a luminous
blue variable \citep{fuc99,eik04,fig05}.\footnote{There has been some 
debate as to the distance to the magnetar, and therefore whether it 
is a member of the star cluster \citep{cam05,mgg05}. However,
such clusters of post-main sequence stars are extremely rare, so 
it seems unlikely that it is a chance association.} 
Red supergiants have not yet
formed, which suggests that the cluster is 3.0--4.5 Myr old, 
and that the progenitor to the magnetar had an initial
mass of $>$50 \msun.
 Likewise, SGR 1900+14 is a member of an anonymous 
star cluster that is probably $<$20 Myr old \citep{vrb00}, which would 
place a lower limit of 
12\msun\ on the initial mass of its progenitor. 
However, further studies of both clusters are warranted, because in 
neither case has the main sequence been identified.

Finally, the AXP 1E 1048.1--5937 lies near the center of a shell of 
neutral hydrogen emission, which resembles the wind-blown bubbles 
produced by 30--40 \msun\ stars \citep{gae05}. There is debate
as to whether the shell and the pulsar are at the same distance
\citep{dvk06}, but if they are associated, than this AXP would also
have had a very massive progenitor.

\section{Massive Stars and Magnetars\label{sec:coeval}}

If further evidence accumulates that magnetars appear have $>$30 \msun\ 
progenitors, it would have implications for both the Galactic population 
of these unusual neutron stars, and for the evolution and deaths of massive
stars. First, if magnetars only descend from massive progenitors, 
then they would be expected to be rare compared to radio pulsars 
\citep{gae05}. 
For a standard initial mass function with $N(>M) \propto M^{-1.3}$ 
\citep{krou02},
only 20\% of stars with $M$$>$8 \msun\ have $M$$>$25 \msun. If half
of the massive stars produce black holes and half produce magnetars,
it would explain why the birth rates of magnetars are only $\sim$10\% 
of those of radio pulsars \citep{kou94,gae99,gae05}.

Second, it implies that some mechanism allows $>$30 \msun\ stars
to lose $>$95\% of their mass and leave behind neutron stars.
This could occur while the stars evolve, either through powerful stellar
winds produced by stars with at least solar metallicity \citep{heg03}, 
or through mass transfer onto a binary companion \citep{wl99}. 
If the mass loss is caused by stellar winds, then one
might expect that the majority of $>$30 \msun\ stars interior to 
the Solar Circle will form neutron stars instead of black holes (and
one might be puzzled by the presence of SGR 0525--66 in the metal-poor 
LMC). If the binary hypothesis is
correct, then the extreme mass loss required to turn a $>$30 \msun\ star
into a neutron star should occur infrequently. Alternatively, some
process during the supernova, such as a strong magnetically-driven wind
\citep{aw05}, might cause the extreme mass loss. In this case, one must
understand how supernovae are triggered before a prediction can be 
made about the number of magnetars.

Finally, the fact that several magnetars had $>$30 \msun\
progenitors suggests that there could be a link between massive stars
and the presence of strong magnetic fields in their remnants.
However, it is not clear when the strong fields of magnetars are
generated.  One possibility is that they are produced by a dynamo that
forms when the core of a rapidly-rotating star collapses
\citep{td93}. In this case, a link with massive stars could result 
if they leave cores that rotate more rapidly, possibly because they 
evolve so rapidly through the red supergiant phase that there is too 
little time for angular momentum to be lost to the expanding envelope 
of the star \citep{heg05}. This scenario would predict that magnetars 
are produced by the majority of massive stars. Another is that 
the $>$$10^{14}$ G 
fields are simply those of the progenitor, which were compressed down 
to the volume of a neutron star
during the collapse that ended the star's life \citep{fw06}. This
hypothesis is motivated by the realization that at least three OB
stars, $\theta^1$ Ori C, HD 191612, and $\tau$ Sco, are
highly-magnetized ($\sim$1 kG) \citep{don02,don06a,don06b}. In this case,
one would expect that the birth rate of magnetars would be proportional
to the fraction of OB stars that are highly-magnetized.

\section{Making Further Progress}

Currently, the magnetars identified in star clusters
provide the most reliable constraints on the initial masses of stars 
that produce neutron stars. However, this sample is small, largely 
because the sample of known Galactic star clusters is 
based on optical observations, and so interstellar extinction makes the 
sample incomplete beyond $\sim$1 kpc of Earth. Therefore, we (PI: D. Figer) 
are performing infrared spectroscopy of candidate star clusters taken from 
the 2MASS \citep{db01,dut03} and {\it Spitzer}/GLIMPSE surveys 
\citep{mer05}, in order to identify more distant, massive clusters that are 
likely to harbor compact objects \citep{fig06}. As the massive clusters
are identified, we are searching them in the radio for pulsars, 
and in X-rays for young, cooling neutron stars and compact objects accreting
from the winds of OB companions.
With a larger sample of compact objects in star clusters, we will
be able to address better which stars produce black holes and neutron stars,
and why.




\begin{theacknowledgments}
I would like to thank J. S. Clark, S. Eikenberry, D. Figer, and 
B. Gaensler for many discussions, which served as the basis of this
review. 
\end{theacknowledgments}


\bibliographystyle{aipproc}   

\end{document}